\documentclass[journal,10pt,twocolumn,final]{IEEEtran}

\ifCLASSOPTIONonecolumn
\renewcommand{\baselinestretch}{1.4}\small\normalsize
\fi
\usepackage{graphics}
\usepackage{rotating}
\usepackage{amsfonts}
\usepackage{amsmath}
\usepackage{subfigure}
\usepackage{multirow}
\usepackage{comment}
\allowdisplaybreaks[1]
\newtheorem{theorem}{Theorem}
\renewcommand{\IEEEQED}{\IEEEQEDopen}

\graphicspath{{Figures/}} \DeclareGraphicsExtensions{.eps,.ps}

\begin{document}

\title{An Upper Bound to the Marginal PDF of the Ordered Eigenvalues of Wishart Matrices$^{\dagger}$}
\ifCLASSOPTIONonecolumn
\small 
\author{\small\IEEEauthorblockN{Hong Ju Park,~\IEEEmembership{\small Member,~IEEE,} and Ender Ayanoglu,~\IEEEmembership{\small Fellow,~IEEE}}\\
\IEEEauthorblockA{Center for Pervasive Communications and Computing\\
Department of Electrical Engineering and Computer Science\\
University of California, Irvine\\
Irvine, CA 92697-2625}
\thanks{$^\dagger$ This letter was presented in part during the IEEE International Conference on Communications, Cape Town, South Africa, May 2010.}}
\vspace{-5mm}
\else
\author{\IEEEauthorblockN{Hong Ju Park,~\IEEEmembership{Member,~IEEE,} and Ender Ayanoglu,~\IEEEmembership{Fellow,~IEEE}\\[-10cm]}
\thanks{$^\dagger$
Hong Ju Park was and Ender Ayanoglu is with the Center for Pervasive Communications and Computing, Department of Electrical
Engineering and Computer Science, University of California, Irvine. Hong Ju Park is now with Samsung Electronics, Seoul, South Korea.
This letter was presented in part during the IEEE International Conference on Communications, Cape Town, South Africa, May 2010.}}
\fi
\maketitle
\normalsize

\ifCLASSOPTIONonecolumn
 \setlength\arraycolsep{4pt}
\else
 \setlength\arraycolsep{2pt}
\fi
\begin{abstract}
Diversity analysis of a number of Multiple-Input Multiple-Output
(MIMO) applications requires the calculation of the expectation of
a function whose variables are the ordered multiple eigenvalues of
a Wishart matrix. In order to carry out this calculation, we need
the marginal pdf of an arbitrary subset of the ordered eigenvalues.
In this letter, we derive an upper
bound to the marginal pdf of the eigenvalues. The derivation is
based on the multiple integration of the well-known joint pdf, which
is very complicated due to the exponential factors of the joint pdf.
We suggest an alternative function that provides simpler calculation
of the multiple integration. As a result, the marginal pdf is shown
to be bounded by a multivariate polynomial with a given degree.
After a standard bounding procedure in a Pairwise Error Probability (PEP) analysis, by
applying the marginal pdf to the calculation of the expectation, the
diversity order for a number of MIMO systems can be obtained in a simple manner.
Simulation results that support the analysis are presented.
\end{abstract} 
\vspace{-4mm}
\section{Introduction} \label{sec:introduction}

In wireless communications, the Wishart matrix arises from the MIMO
transmission environment, where the channel matrix is modeled as
complex Gaussian, as in the Rayleigh fading model
\cite{edelmanThesis}. In particular, if the channel matrix is
available at the transmitter as well as at the receiver, the
beamforming matrices can be obtained from the singular value
decomposition (SVD) of the channel matrix to build the diagonalizing
structure, known to be optimal to maximize the performance
\cite{palomarTSP03}. In the {\em uncoded\/} version of
this multiple beamforming scheme, the
diversity order, an important performance measure of MIMO systems at
the high signal-to-noise ratio regime, is determined by the
subchannel with the smallest eigenvalue of the Wishart
matrix \cite{sengulTC06AnalSingleMultpleBeam}, \cite{OrdonezTSP07},
\cite{ZanellaGlobe09}. In general, the diversity order is calculated
from the PEP, expressed as $E
\left[ \kappa (\mu_l) \right]$, where $\kappa(\mu_l)$ is a function
of the $l^{th}$ ordered eigenvalue $\mu_l$, and $E[\,\cdot\,]$ is the
expectation operator. In order to carry out this calculation, we need
to find the marginal pdf
of the single eigenvalue $\mu_l$ of the Wishart matrix. A first
order polynomial expansion is used to derive the simple closed form
expression of the marginal pdf in \cite{OrdonezTSP07} and
\cite{KhoshnevisProcAllerton04}, while the more accurate expression
as the sums of terms of the form $\mu_l^x
e^{y\mu_l}$ is provided in \cite{ZanellaGlobe09}. The resulting
diversity order of multiple beamforming with the $l^{th}$ eigenvalue
involved is $(N - l + 1)(M - l +1)$ where $N$ and $M$ are the number
of transmit and receive antennas, respectively
\cite{sengulTC06AnalSingleMultpleBeam}, \cite{OrdonezTSP07}.

The average PEP between two codewords in the {\em coded\/} multiple
beamforming scheme, on the other hand, requires the calculation of
the expectation $E \left[ \phi(\mu_1, \cdots, \mu_Y ) \right]$,
where $\phi(\cdot)$ is a function with multiple ordered eigenvalues
involved \cite{akayTC06BICMB}. For the pairwise codewords whose
corresponding function $\phi(\cdot)$ includes all of the singular
values available from the SVD of the channel matrix, authors in
\cite{akayTC06BICMB} calculated the diversity order from the simple
closed form expression of the average PEP, by making use of the fact
that the sum of all ordered eigenvalues follows a chi-squared
distribution. If $\phi(\cdot)$ is composed of a subset of the
ordered eigenvalues, the calculation of the expectation needs the
marginal pdf of the eigenvalues. The closed form expressions of
consecutive and an arbitrary subset of ordered eigenvalues are given
in 
\cite{ZanellaJCOM09},
while the expressions for unordered eigenvalues are provided in
\cite{MaarefJWCOM07} and \cite{MaarefJWCOM07_2}.
A difficulty exists in determining an analytical
diversity figure with these prior approaches. They are typically
in the form of a product of integrals to be calculated, and consist of
the incomplete Gamma functions that enable numerical evaluation, but
make the analysis difficult.

In this letter, we propose a methodology to calculate an upper bound
to the marginal pdf of the ordered eigenvalues. Then, we derive the
diversity order by using the upper bound to the marginal pdf. Since
the direct calculation of the marginal pdf is very complicated due
to the multiple integration of the joint pdf which has the
exponential function, we suggest an alternative function as a
substitute for the joint pdf to simplify the multiple integration.
The resulting diversity order is $(N - p_1 + 1)(M - p_1 + 1)$ where
$p_1$ is the index to indicate the best among the eigenvalues
appearing in the $\phi(\cdot)$ function.

\vspace{-2mm}
\section{Problem Statement} \label{sec:statement}

The elements of the MIMO channel $\mathbf{H} \in \mathbb{C}^{M
\times N}$ are assumed to be Gaussian with zero mean and unit
variance. In addition, the covariance matrix, which is defined as
$\mathbf{V}_j = E [ \mathbf{h}_j \mathbf{h}_j^\dag ]$ where
$\mathbf{h}_j$ is the $j^{th}$ column vector of $\mathbf{H}$, and
$\dag$ stands for conjugate transpose, satisfies $\mathbf{V}_j =
\mathbf{I}$ for all $j$. Based on the assumption above, the matrix
$\mathbf{HH}^{\dag}$ is called uncorrelated central Wishart matrix
\cite{ZanellaGlobe09}. The $i^{th}$ eigenvalue of
$\mathbf{HH}^{\dag}$, denoted by $\mu_i$, is sorted such that $\mu_i
> \mu_j$ for $i < j$. Throughout this letter, we use $X$ and $Y$ as $X = \max ( N,
M )$, and $Y = \min ( N, M )$.

The average pairwise error probability that the receiver decides
$\mathbf{\hat{c}}$ instead of $\mathbf{c}$ as the transmitted signal
is upper bounded by \cite{akayTC06BICMB}
\begin{align}
\mathrm{Pr}\left(\mathbf{c} \rightarrow \mathbf{\hat{c}}\right) \leq
E \left[ \exp \left(- \gamma \sum\limits_{j=1}^{Y} \alpha_j \mu_{j}
\right) \right] \label{eq:PEP_expression}
\end{align}
where $\gamma$ is the signal-to-noise ratio, and $\alpha_j$ is a
given non-negative real value. We note that a bound of this form can be obtained
for a number of MIMO SVD systems, e.g., \cite{ParkICC09}, \cite{ParkGlobecom09}, \cite{ParkICC10}, \cite{LiGlobecom10}.
Let's define $\alpha_{min}$ as the
minimum among the nonzero $\alpha$ values. Using the inequality
$\sum_{j=1}^{Y} \alpha_{j} \mu_{j} \geq \alpha_{min}
\sum_{j=1,\alpha_j \neq 0}^{Y} \mu_{j}$, we rewrite
(\ref{eq:PEP_expression}) as
\begin{align}
\mathrm{Pr}\left(\mathbf{c} \rightarrow \mathbf{\hat{c}}\right) \leq
E \left[ \exp \left(- \gamma \alpha_{min} \sum\limits_{k=1}^{K}
\mu_{p_k} \right) \right] \label{eq:PEP_min_expression}
\end{align}
where $p_k$ is the $k^{th}$ element of a vector $\mathbf{p} = [p_1
\, \cdots \, p_K]^T$ whose elements are the indices corresponding to
non-zero $\alpha$, i.e., \mbox{$\alpha_{p_k} \neq 0$}. Similarly,
$\mathbf{s} = [s_1 \, \cdots \, s_{(Y-K)}]^T$ is defined as a vector
whose elements are the indices $k$ such that $\alpha_{s_k} = 0$. The
vectors $\mathbf{p}$ and $\mathbf{s}$ are sorted in increasing
order. To calculate (\ref{eq:PEP_min_expression}), we need the
marginal pdf of the $K$ eigenvalues by calculating the multiple
integration over the domain $\mathcal{D}_{\mathbf{s}}$
\ifCLASSOPTIONonecolumn
\begin{align}
f \left( \mu_{p_1}, \cdots, \mu_{p_K} \right ) = \int \cdots
\int_{\mathcal{D}_{\mathbf{s}}} \rho \left( \mu_1, \cdots, \mu_Y
\right) \, d \mu_{s_{(Y-K)}} \, \cdots \, d \mu_{s_1}.
\label{eq:Marginal_PDF}
\end{align}
\else
\begin{multline}
f \left( \mu_{p_1}, \cdots, \mu_{p_K} \right ) = \\
\int \cdots \int_{\mathcal{D}_{\mathbf{s}}} \rho \left( \mu_1,
\cdots, \mu_Y \right) \, d \mu_{s_{(Y-K)}} \, \cdots \, d \mu_{s_1}.
\label{eq:Marginal_PDF}
\end{multline}
\fi The joint pdf of the ordered strictly positive eigenvalues of
the uncorrelated central Wishart matrices $\rho \left(\mu_1, \cdots,
\mu_Y \right)$ in (\ref{eq:Marginal_PDF}) is available in the
literature \cite{edelmanThesis}, \cite{Verdubook} as
\begin{align}
\rho \left(\mu_1, \cdots, \mu_Y \right) = \psi\left(\mu_1, \cdots,
\mu_Y \right) e^{-\sum\limits_{j=1}^Y \mu_j} \label{eq:Joint_PDF}
\end{align}
where the polynomial $\psi\left(\mu_1, \cdots, \mu_Y \right)$ is
\begin{align}
\psi\left(\mu_1, \cdots, \mu_Y \right) = \prod^Y_{i=1} \mu_{i}^{X-Y}
\prod^Y_{j>i} \left (\mu_i - \mu_j \right )^2.
\label{eq:Polynomial_p}
\end{align}
Because we are interested in the exponent of $\gamma$, a constant multiplier,
which appears in the literature, and is irrelevant to the exponent
of $\gamma$, is ignored in (\ref{eq:Polynomial_p}) for brevity.

The closed form expression of the marginal pdf can be calculated by
evaluating (\ref{eq:Marginal_PDF}). However, this evaluation is
complicated due to the multiple integration of the product of the
polynomial and the exponential function in
(\ref{eq:Marginal_PDF})-(\ref{eq:Polynomial_p}). Alternatively, we
will now develop a method to get a simple expression for an upper
bound to the marginal pdf. Then, we will use the upper bound to
calculate (\ref{eq:PEP_min_expression}).

\vspace{-2mm}
\section{An Upper Bound to the Marginal PDF} \label{sec:derivation}

The complexity of the multiple integration to calculate the marginal
pdf in (\ref{eq:Marginal_PDF}) mainly comes from the fact that the
elementary integration inside the multiple integration, $\int_0^x
y^m e^{-y} dy$, generates a large number of terms of the form $x^n
e^{-x}$ for large $m$, i.e.,
\ifCLASSOPTIONonecolumn
\begin{align}
\int_0^xy^me^{-y}dy & = \left[-y^me^{-y}\right|_0^{y=x}+m\int_0^xy^{m-1}e^{-y}dy\notag\\
& = -e^{-x}(x^m+mx^{m-1}+m(m-1)x^{m-2}+\cdots+m!)+m! .
\end{align}
\else
\begin{align}
\int_0^xy^me^{-y}dy & = \left[-y^me^{-y}\right|_0^{y=x}+m\int_0^xy^{m-1}e^{-y}dy\notag\\
& = -e^{-x}(x^m+mx^{m-1}\\
& \qquad +m(m-1)x^{m-2}+\cdots+m!)+m! .\notag
\end{align}
\fi
However, if we remove the exponential
function from the elementary integration, the integration produces
only one term, resulting in a much simpler multiple integration. In
addition, since the eigenvalues of the Wishart matrix are positive
and real, \mbox{$e^{-\mu_i} \leq 1$} holds true for any $i$. This
idea leads to a simple result of the elementary integration as
\begin{align}
\int_0^x y^m e^{-y} dy \leq \frac{1}{m+1}x^{m+1}.
\end{align}

To apply the idea above to the calculation of the marginal pdf, we
introduce an alternative function
\ifCLASSOPTIONonecolumn
\begin{align}
\hat{\rho} \left(\mu_1, \cdots, \mu_Y \right) = \left\{
\begin{array}{ll}
\psi\left(\mu_1, \cdots, \mu_Y \right) e^{-\left(\mu_1 +
\sum\limits_{k=1}^K \mu_{p_k}\right)}
& \textrm{if $\alpha_1 = 0$} \\
\psi\left(\mu_1, \cdots, \mu_Y \right) e^{-\sum\limits_{k=1}^K
\mu_{p_k}} & \textrm{if $\alpha_1 > 0$}
\end{array}
\right. \label{eq:Joint_PDF_2}
\end{align}
\else
\begin{multline}
\hat{\rho} \left(\mu_1, \cdots, \mu_Y \right) = \\
\left\{
\begin{array}{ll}
\psi\left(\mu_1, \cdots, \mu_Y \right) e^{-\left(\mu_1 + \sum\limits_{k=1}^K \mu_{p_k}\right)} \\
& \textrm{if $\alpha_1 = 0$} \\
\psi\left(\mu_1, \cdots, \mu_Y \right) e^{-\sum\limits_{k=1}^K
\mu_{p_k}} & \textrm{if $\alpha_1 > 0$}
\end{array}
\right. \label{eq:Joint_PDF_2}
\end{multline}
\fi where the exponential factors irrelevant to the variables of
integration are removed, except $\mu_1$ which is kept in the case of
$\alpha_1 = 0$. The reason for keeping $\mu_1$ will be explained
later. Correspondingly, let's define $\hat{f} \left( \mu_{p_1},
\cdots, \mu_{p_K} \right )$ in a similar fashion to
(\ref{eq:Marginal_PDF}) with $\rho \left( \mu_1, \cdots, \mu_Y
\right)$ replaced by \mbox{$\hat{\rho} \left( \mu_1, \cdots, \mu_Y
\right)$}, that is, \ifCLASSOPTIONonecolumn
\begin{align}
\hat{f} \left( \mu_{p_1}, \cdots, \mu_{p_K} \right ) = \int \cdots
\int_{\mathcal{D}_{\mathbf{s}}} \hat{\rho} \left(\mu_1, \cdots,
\mu_Y \right) d \mu_{s_{(Y-K)}} \, \cdots \, d \mu_{s_1}.
\label{eq:Marginal_PDF_2}
\end{align}
\else
\begin{multline}
\hat{f} \left( \mu_{p_1}, \cdots, \mu_{p_K} \right ) = \\
\int \cdots \int_{\mathcal{D}_{\mathbf{s}}} \hat{\rho} \left(\mu_1,
\cdots, \mu_Y \right) d \mu_{s_{(Y-K)}} \, \cdots \, d \mu_{s_1}.
\label{eq:Marginal_PDF_2}
\end{multline}
\fi We see that $\rho \left(\mu_1, \cdots, \mu_Y \right) \leq
\hat{\rho} \left(\mu_1, \cdots, \mu_Y \right)$ for either the  case of
$\alpha_1 = 0$ or $\alpha_1 > 0$, and therefore, $f \left(
\mu_{p_1}, \cdots, \mu_{p_K} \right ) \leq \hat{f} \left( \mu_{p_1},
\cdots, \mu_{p_K} \right )$, where $\hat{f} \left( \mu_{p_1},
\cdots, \mu_{p_K} \right )$ has a simpler expression. We will
employ (\ref{eq:Marginal_PDF_2}) to calculate our bound in the next
subsections.

\subsubsection{For $\alpha_1 = 0$} \label{sec:subsec_1}

Since some factors of $\hat{\rho}\left( \mu_{1}, \cdots, \mu_{Y} \right)$
are irrelevant to the variables of integration, they can be moved
out. By defining a polynomial $g \left( \mu_{p_1}, \cdots, \mu_{p_K}
\right)$
\begin{align}
g \left( \mu_{p_1}, \cdots, \mu_{p_K} \right) = \prod^K_{k=1}
\mu_{p_k}^{X-Y} \prod^K_{j>k} \left( \mu_{p_k} - \mu_{p_j} \right
)^2, \label{eq:function_g}
\end{align}
we rewrite (\ref{eq:Marginal_PDF_2}) as
\begin{multline}
\hat{f} \left( \mu_{p_1}, \cdots, \mu_{p_K} \right ) = g\left(
\mu_{p_1}, \cdots, \mu_{p_K} \right )
e^{-\sum\limits_{k=1}^K \mu_{p_k}} \\
\times \int_0^{\infty} e^{-\mu_1} \int_0^{\mu_{s_2-1}} \cdots
\int_0^{\mu_{s_{(Y-K)}-1}} \ifCLASSOPTIONtwocolumn \\ \times \fi
\frac{\psi\left( \mu_{1}, \cdots, \mu_{Y} \right)}{g\left(
\mu_{p_1}, \cdots, \mu_{p_K} \right)} d\mu_{s_{(Y-K)}} \cdots
d\mu_{s_2} d\mu_{1} \label{eq:altMarginal_PDF_decomposed_1}
\end{multline}
where $\frac{\psi\left( \mu_{1}, \cdots, \mu_{Y} \right)}{g\left(
\mu_{p_1}, \cdots, \mu_{p_K} \right)}$ is a polynomial of the
remained factors, and $d\mu_{s_1}$ is replaced by $d\mu_1$ since
$s_1 = 1$ for $\alpha_1 = 0$. The multiple integration of
$\frac{\psi\left( \mu_{1}, \cdots, \mu_{Y} \right)}{g\left(
\mu_{p_1}, \cdots, \mu_{p_K} \right)}$ over the variables
$\mu_{s_{(Y-K)}}$, $\cdots$, $\mu_{s_2}$ results in an intermediate
polynomial whose terms are composed of the variables $\mu_1$,
$\mu_{p_1}$, $\cdots$, $\mu_{p_K}$. In other words, the intermediate
polynomial is the sum of the terms of the form as $\mu_{p_1}^{x_1}
\times \cdots \times \mu_{p_K}^{x_K} \int_0^\infty \mu_1^y
e^{-\mu_1} d \mu_1$. The final integration of the intermediate
polynomial over $\mu_1$ leaves a polynomial with the variables
$\mu_{p_1}$, $\cdots$, $\mu_{p_K}$ since $\int_0^\infty \mu_1^y
e^{-\mu_1} d \mu_1 = y!$. It should now be clear why we keep
$e^{-\mu_1}$ in $\hat{\rho} \left( \mu_1, \cdots, \mu_Y \right)$.
The absence of this factor would have led the integration to
diverge.

Defining $h_1 \left( \mu_{p_1}, \cdots, \mu_{p_K} \right)$ as the
result of the multiple integration, we rewrite
(\ref{eq:altMarginal_PDF_decomposed_1}) as
\begin{align}
\hat{f} \left( \mu_{p_1}, \cdots, \mu_{p_K} \right) = r_1 \left(
\mu_{p_1}, \cdots, \mu_{p_K} \right) e^{-\sum\limits_{k=1}^K
\mu_{p_k}} \label{eq:upperbound_f}
\end{align}
where the polynomial $r_1 \left( \mu_{p_1}, \cdots, \mu_{p_K}
\right)$ is defined as \ifCLASSOPTIONonecolumn
\begin{align}
r_1 \left( \mu_{p_1}, \cdots, \mu_{p_K} \right) = h_1
\left(\mu_{p_1}, \cdots, \mu_{p_K} \right ) \times g\left(
\mu_{p_1}, \cdots, \mu_{p_K} \right ). \label{eq:function_r1}
\end{align}
\else
\begin{multline}
r_1 \left( \mu_{p_1}, \cdots, \mu_{p_K} \right) = \\
h_1 \left(\mu_{p_1}, \cdots, \mu_{p_K} \right ) \times g\left(
\mu_{p_1}, \cdots, \mu_{p_K} \right ). \label{eq:function_r1}
\end{multline}
\fi

\subsubsection{For $\alpha_1 > 0$} \label{sec:subsec_2}

Using the polynomial $g \left( \mu_{p_1}, \cdots, \mu_{p_K} \right)$
in (\ref{eq:function_g}), we rewrite (\ref{eq:Marginal_PDF_2}) in
this case as \ifCLASSOPTIONonecolumn
\begin{multline}
\hat{f} \left( \mu_{p_1}, \cdots, \mu_{p_K} \right ) = g\left(
\mu_{p_1}, \cdots, \mu_{p_K} \right )
e^{-\sum\limits_{k=1}^K \mu_{p_k}} \\
\times \int_0^{\mu_{s_1-1}} \int_0^{\mu_{s_2-1}} \cdots
\int_0^{\mu_{s_{(Y-K)}-1}} \frac{\psi\left( \mu_{1}, \cdots, \mu_{Y}
\right)}{g\left( \mu_{p_1}, \cdots, \mu_{p_K} \right)}
d\mu_{s_{(Y-K)}} \cdots d\mu_{s_2} d\mu_{s_1}
\label{eq:altMarginal_PDF_decomposed_2}
\end{multline}
\else
\begin{multline}
\hat{f} \left( \mu_{p_1}, \cdots, \mu_{p_K} \right ) = g\left(
\mu_{p_1}, \cdots, \mu_{p_K} \right )
e^{-\sum\limits_{k=1}^K \mu_{p_k}} \\
\times \int_0^{\mu_{s_1-1}} \int_0^{\mu_{s_2-1}} \cdots
\int_0^{\mu_{s_{(Y-K)}-1}} \\
\times \frac{\psi\left( \mu_{1}, \cdots, \mu_{Y} \right)}{g\left(
\mu_{p_1}, \cdots, \mu_{p_K} \right)} d\mu_{s_{(Y-K)}} \cdots
d\mu_{s_2} d\mu_{s_1} \label{eq:altMarginal_PDF_decomposed_2}
\end{multline}
\fi where $\frac{\psi\left( \mu_{1}, \cdots, \mu_{Y}
\right)}{g\left( \mu_{p_1}, \cdots, \mu_{p_K} \right)}$ is the same
kind of polynomial that is described in the previous subsection. The
multiple integration results in a polynomial of the variables
$\mu_{p_1}$, $\cdots$, $\mu_{p_K}$ without diverging since the last
integration is not involved with infinity. By defining $h_2 \left(
\mu_{p_1}, \cdots, \mu_{p_K} \right)$ as the result of the multiple
integration of (\ref{eq:altMarginal_PDF_decomposed_2}), we get
$\hat{f} \left( \mu_{p_1}, \cdots, \mu_{p_K} \right)$ as
\begin{align}
\hat{f} \left( \mu_{p_1}, \cdots, \mu_{p_K} \right) = r_2 \left(
\mu_{p_1}, \cdots, \mu_{p_K} \right) e^{-\sum\limits_{k=1}^K
\mu_{p_k}} \label{eq:upperbound_f2}
\end{align}
where $r_2 \left( \mu_{p_1}, \cdots, \mu_{p_K} \right)$ is defined
as \ifCLASSOPTIONonecolumn
\begin{align}
r_2 \left( \mu_{p_1}, \cdots, \mu_{p_K} \right) = h_2
\left(\mu_{p_1}, \cdots, \mu_{p_K} \right ) \times g\left(
\mu_{p_1}, \cdots, \mu_{p_K} \right ). \label{eq:function_r2}
\end{align}
\else
\begin{multline}
r_2 \left( \mu_{p_1}, \cdots, \mu_{p_K} \right) = \\
h_2 \left(\mu_{p_1}, \cdots, \mu_{p_K} \right ) \times g\left(
\mu_{p_1}, \cdots, \mu_{p_K} \right ). \label{eq:function_r2}
\end{multline}
\fi


The polynomials $r_1 \left( \mu_{p_1}, \cdots, \mu_{p_K} \right)$
and $r_2 \left( \mu_{p_1}, \cdots, \mu_{p_K} \right)$ are
multivariate polynomials with many terms. It is worthwhile to focus
on the smallest degree of the terms because it plays an important
role in determining the behavior of (\ref{eq:PEP_expression}) in the
high signal-to-noise ratio regime. The mathematical analysis of the
behavior will be described in Section \ref{sec:calculation} with the
help of the next Theorem.

\begin{theorem}
The smallest degree of the multivariate polynomial $r_1 \left(
\mu_{p_1}, \cdots, \mu_{p_K} \right)$ or $r_2 \left( \mu_{p_1},
\cdots, \mu_{p_K} \right)$ is $(N - p_1 + 1)(M -p_1 + 1) - K$.
\label{theorem:smallest_degree}
\end{theorem}
\begin{IEEEproof}
See Appendix \ref{sec:appendix_A}.
\end{IEEEproof}

In the case of the single eigenvalue where $K = 1$, and $p_1 = l$,
Theorem \ref{theorem:smallest_degree} states that the smallest
degree of $\mu_l$ is $(N-l+1)(M-l+1)-1$. This generalizes the result
of the first order expansion in \cite{OrdonezTSP07},
\cite{KhoshnevisProcAllerton04} to calculate the marginal pdf of the
$l^{th}$ eigenvalue.

\vspace{-4mm}
\section{Calculation of the Expectation} \label{sec:calculation}

According to the analysis in Section \ref{sec:derivation}, the
marginal pdf $f \left( \mu_{p_1}, \cdots, \mu_{p_K} \right)$ is
upper bounded by the general expression
\begin{align*}
f \left( \mu_{p_1}, \cdots, \mu_{p_K} \right) \leq r \left(
\mu_{p_1}, \cdots, \mu_{p_K} \right) e^{-\sum\limits_{k=1}^K
\mu_{p_k}}
\end{align*}
where $r \left( \mu_{p_1}, \cdots, \mu_{p_K} \right)$ is a
polynomial with the smallest degree of $(N - p_1 + 1)(M -p_1 + 1) -
K$. We are now ready to obtain an upper bound to
(\ref{eq:PEP_min_expression}) by calculating
\ifCLASSOPTIONtwocolumn
\begin{align}
E &\left[ \exp \left(- \gamma \alpha_{min} \sum\limits_{k=1}^{K}
\mu_{p_k} \right) \right] \nonumber \\ &= \int\cdots
\int_{\mathcal{D}_{\mathbf{p}}} \exp \left(- \gamma \alpha_{min}
\sum\limits_{k=1}^{K} \mu_{p_k} \right) \nonumber \\& \qquad \times
f \left( \mu_{p_1}, \cdots, \mu_{p_K} \right) d \mu_{p_K} \cdots d
\mu_{p_1} \label{eq:PEP_final} \\
&\leq \int_0^{\infty} \cdots \int_0^{\mu_{p_{K-1}}} r \left(
\mu_{p_1}, \cdots, \mu_{p_K} \right ) \nonumber
\\& \qquad \times e^{-(1+\gamma \alpha_{min})
\sum\limits_{k=1}^K \mu_{p_k}}d \mu_{p_K} \cdots d \mu_{p_1}
\nonumber
\end{align}
\else
\begin{multline}
E \left[ \exp \left(- \gamma \alpha_{min} \sum\limits_{k=1}^{K}
\mu_{p_k} \right) \right] = \int\cdots
\int_{\mathcal{D}_{\mathbf{p}}} \exp \left(- \gamma \alpha_{min}
\sum\limits_{k=1}^{K} \mu_{p_k} \right) f \left( \mu_{p_1}, \cdots,
\mu_{p_K} \right) d \mu_{p_K} \cdots d
\mu_{p_1}\\
\leq \int_0^{\infty} \cdots \int_0^{\mu_{p_{K-1}}} r \left(
\mu_{p_1}, \cdots, \mu_{p_K} \right ) e^{-(1+\gamma \alpha_{min})
\sum\limits_{k=1}^K \mu_{p_k}}d \mu_{p_K} \cdots d \mu_{p_1}
\label{eq:PEP_final}
\end{multline}
\fi where $\mathcal{D}_{\mathbf{p}}$ is the domain of integration.
Note that $e^{-(1+\gamma \alpha_{min})\sum_{k=1}^K\mu_{p_k}} <
e^{-\gamma \alpha_{min}\sum_{k=1}^K\mu_{p_k}}$.
In Theorem \ref{theorem:Elemental_Integral}, we provide
the result of the multiple integration of a term whose degree is
$\sum_k \beta_k$.

\begin{theorem}
For a multivariate term with variables $\theta_i$ for $i = 1,
\cdots, K$ whose exponent, denoted by $\beta_i$, is a non-negative
integer, the multiple integration in the domain $\infty > \theta_1
> \theta_2 > \cdots > \theta_K > 0$ is
\ifCLASSOPTIONtwocolumn
\begin{multline}
\int_0^{\infty} \cdots \int_0^{\theta_{K-1}} \theta_1^{\beta_1}
\cdots \theta_K^{\beta_K} e^{-\omega \sum\limits_{k=1}^K \theta_{k}}
d \theta_{K}
\cdots d \theta_{1} = \\
\zeta \omega^{ -\left( K + \sum\limits_{k=1}^K
\beta_k \right)} \label{eq:Exponent_calculation}
\end{multline}
\else
\begin{align}
\int_0^{\infty} \cdots \int_0^{\theta_{K-1}} \theta_1^{\beta_1}
\cdots \theta_K^{\beta_K} e^{-\omega \sum\limits_{k=1}^K \theta_{k}}
d \theta_{K} \cdots d \theta_{1} = \zeta \omega^{ -\left( K +
\sum\limits_{k=1}^K \beta_k \right)} \label{eq:Exponent_calculation}
\end{align}
\fi where $\zeta$ is a constant. \label{theorem:Elemental_Integral}
\end{theorem}
\begin{IEEEproof}
See Appendix \ref{sec:appendix_B}.
\end{IEEEproof}

Since the polynomial $r \left( \mu_{p_1}, \cdots, \mu_{p_K} \right
)$ is the sum of a number of terms with different degrees, the
result of (\ref{eq:PEP_final}) is also the sum of the terms of
$(\gamma \alpha_{min})$ whose exponent obeys Theorem
\ref{theorem:Elemental_Integral}. For large $\gamma$, it is easy to
see that the overall sum is dominated by the term with the smallest
degree of $\gamma^{-1}$. Theorem \ref{theorem:Elemental_Integral}
indicates that the smallest degree of $\gamma^{-1}$ results from the
smallest degree of $r \left( \mu_{p_1}, \cdots, \mu_{p_K} \right )$.
Therefore, we conclude that (\ref{eq:PEP_expression}) is upper
bounded by \ifCLASSOPTIONtwocolumn
\begin{multline}
E \left[ \exp \left(- \gamma \sum\limits_{j=1}^{Y} \alpha_j \mu_{j}
\right) \right] \leq \\
\eta \left( \gamma \alpha_{min} \right)^{-(N-p_1+1)(M-p_1+1)}
\label{eq:PEP_result}
\end{multline}
\else
\begin{align}
E \left[ \exp \left(- \gamma \sum\limits_{j=1}^{Y} \alpha_j \mu_{j}
\right) \right] \leq \eta \left( \gamma \alpha_{min}
\right)^{-(N-p_1+1)(M-p_1+1)} \label{eq:PEP_result}
\end{align}
\fi where $\eta$ is a constant, and irrelevant to $\gamma$.

\vspace{-4mm}
\section{Simulation Results} \label{sec:simulation}

Fig. \ref{fig:3x3} shows the calculation of
(\ref{eq:PEP_expression}) where $N=M=3$, with several specific
$\alpha$ values through a Monte-Carlo simulation. The legend
represents the $\alpha$ values as a vector notation $[\alpha_1
\cdots \alpha_S]$. Three dotted straight lines are the asymptotes
whose exponents correspond to $1$, $4$, and $9$. The curves of the
$\alpha$ values $[1 \,\, 0 \,\, 0]$, $[0.1 \,\, 0 \,\, 1]$, and $[3
\,\, 0 \,\, 5]$, whose $p_1$ is $1$, are parallel to the asymptote
of $(3-1+1)(3-1+1) = 9$. By comparing the slopes of the other curves
with the asymptotes, we see that the analysis is supported by the
simulation.

\ifCLASSOPTIONonecolumn
\begin{figure}[!m]
\centering \includegraphics[width = 0.5\linewidth]{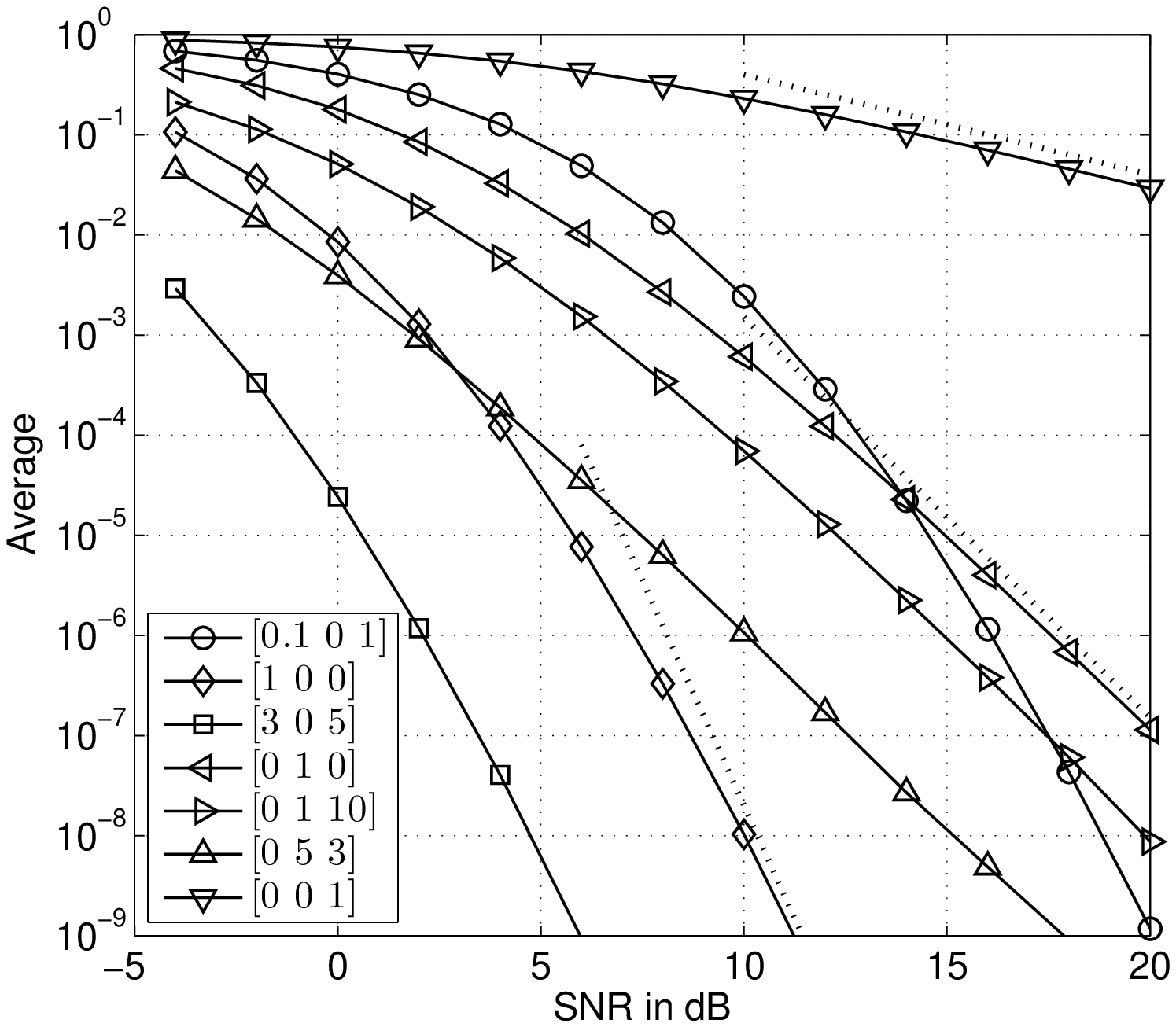}
\caption{Simulation results together with asymptotic diversity orders that show the validity of the technique for $N = M = 3$. The vector $[\alpha_1\;\alpha_2\;\alpha_3]$ is provided in the legend. Note the importance of the first $\alpha_i$ equal to zero in determining the diversity order.}
\label{fig:3x3}
\end{figure}
\else
\begin{figure}[!t]
\centering \includegraphics[width = 1\linewidth]{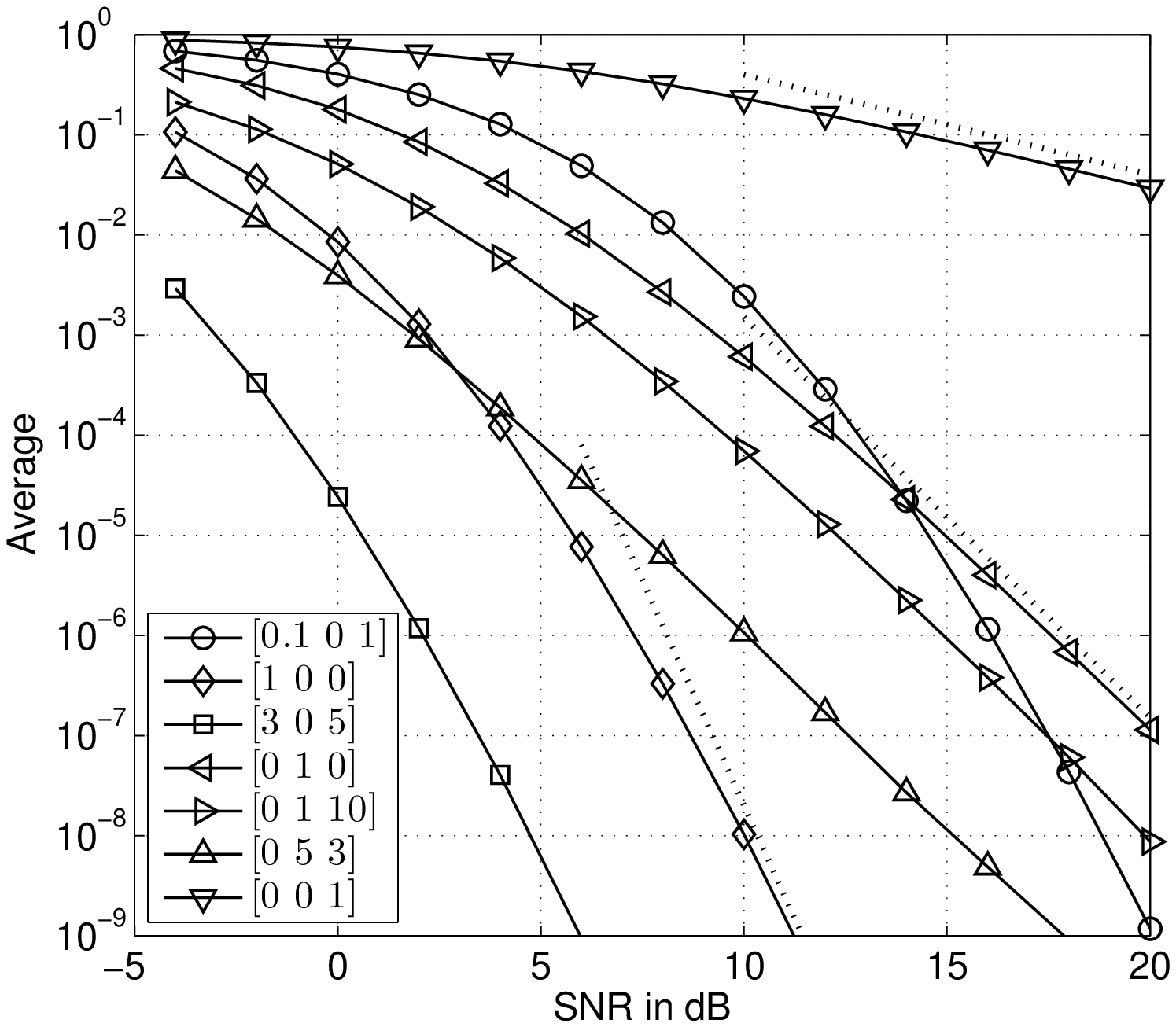}
\caption{Simulation results together with asymptotic diversity orders that show the validity of the technique for $N = M = 3$. The vector $[\alpha_1\;\alpha_2\;\alpha_3]$ is provided in the legend. Note the importance of the first $\alpha_i$ equal to zero in determining the diversity order.} \label{fig:3x3}
\end{figure}
\fi

Fig. \ref{fig:4x4} depicts the simulation result of $N=M=4$. The
dotted lines are the asymptotes of the exponents $4$, $9$ and $16$.
A comparison of a slope with the asymptote reveals that simulation
matches the analysis. Note that even though $\alpha$ corresponding
to the smallest eigenvalue is $100$ times that of the best
eigenvalue, the slope is determined by the best eigenvalue.

\ifCLASSOPTIONonecolumn
\begin{figure}[!m]
\centering \includegraphics[width = 0.5\linewidth]{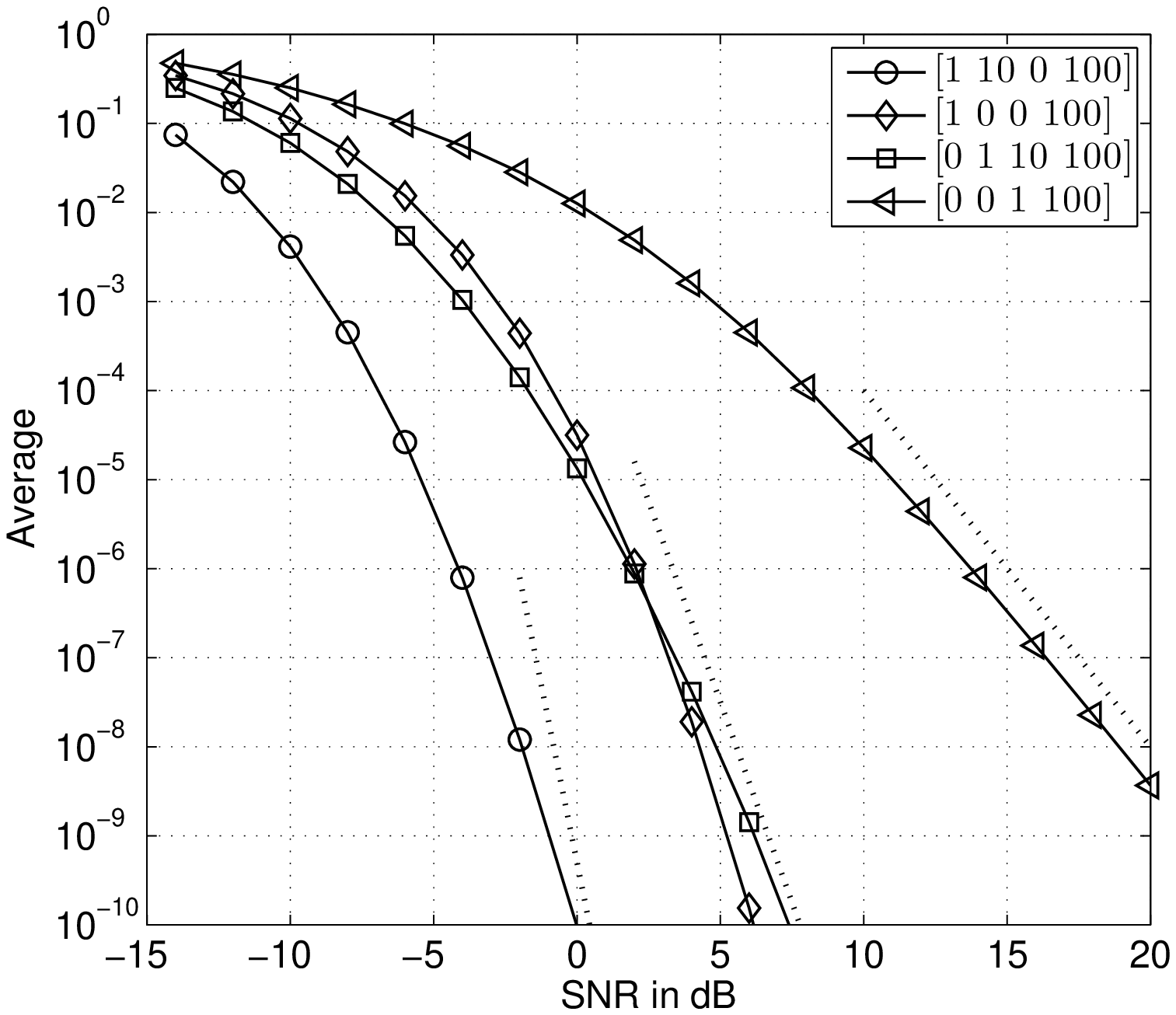}
\caption{Simulation results together with asymptotic diversity orders that show the validity of the technique for $N = M = 4$. Note the accuracy of the technique even if the first nonzero eigenvalue may be much smaller than the successive ones.} \label{fig:4x4}
\end{figure}
\else
\begin{figure}[!t]
\centering \includegraphics[width = 1\linewidth]{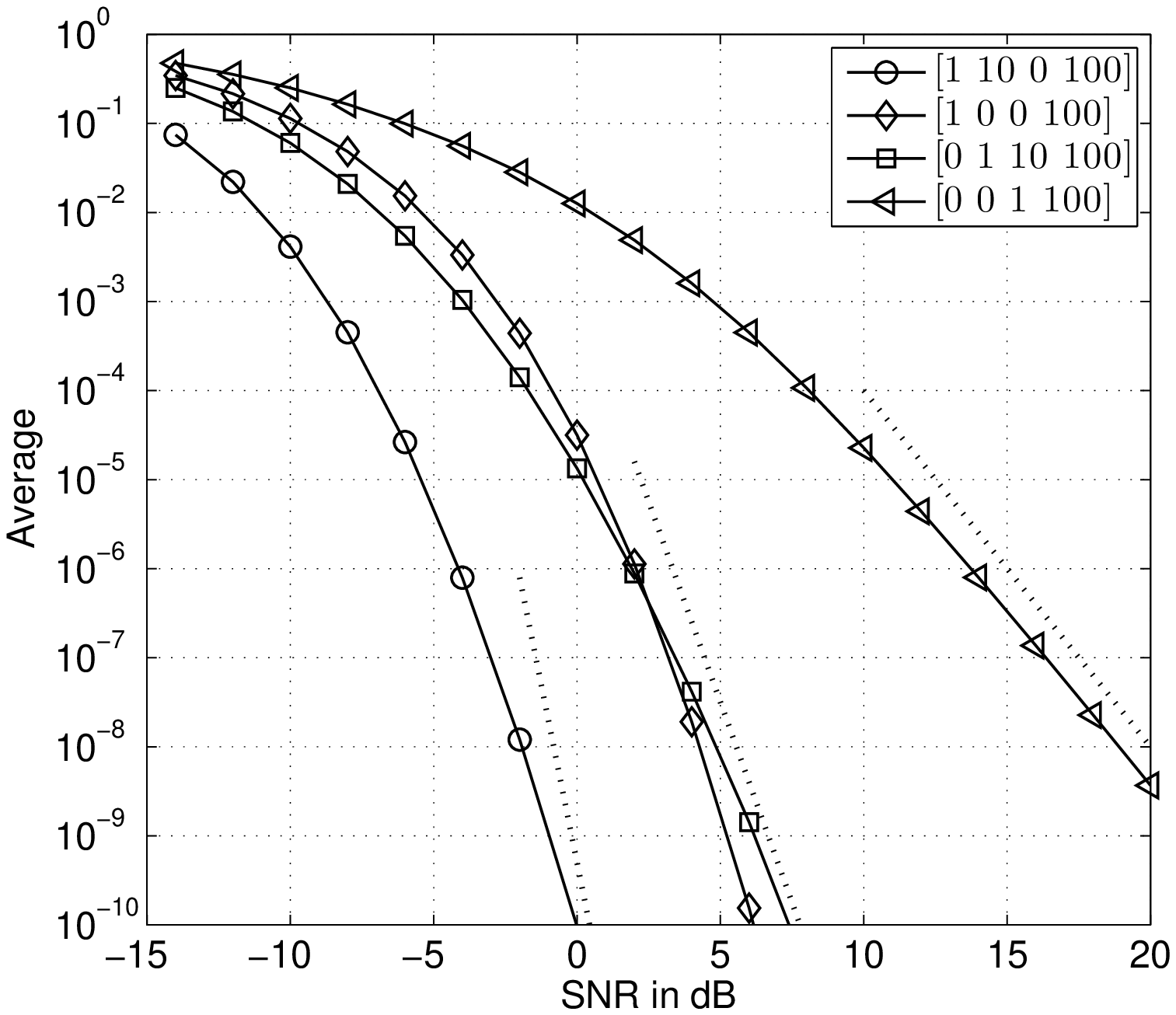}
\caption{Simulation results together with asymptotic diversity orders that show the validity of the technique for $N = M = 4$. Note the accuracy of the technique even if the first nonzero eigenvalue may be much smaller than the successive ones.} \label{fig:4x4}
\end{figure}
\fi

\vspace{-4mm}
\section{Conclusion} \label{sec:conclusion}

We derived an upper bound to the marginal pdf of the ordered
eigenvalues of a complex central Wishart matrix. Such matrices arise in the analysis of MIMO SVD systems.
Our bound employs an alternative function to simplify the multiple integration of the pdf.

For MIMO systems employing SVD, a standard bounding technique provides a simple bound such as (\ref{eq:PEP_expression}),
see e.g., \cite{sengulTC06AnalSingleMultpleBeam},
\cite{akayTC06BICMB},\cite{ParkICC09}, \cite{ParkGlobecom09}, \cite{ParkICC10}, \cite{LiGlobecom10}.
By applying our
result to the calculation of the expectation, one can calculate the
diversity order $(N - p_1 + 1)(M - p_1 + 1)$ in the high
signal-to-noise ratio regime for a system
where $N$ and $M$ are the number of transmit and receive antennas, respectively,
and $p_1$ is the index to the smallest nonzero weight value in a PEP analysis. Simulation results provided here and elsewhere
support the validity of the technique for a variety of MIMO SVD systems with a number of different parameters, see, e.g., \cite{ParkICC09}, \cite{ParkGlobecom09}, \cite{ParkICC10}, \cite{LiGlobecom10}.


\appendices
\vspace{-2mm}
\section{Proof of Theorem \ref{theorem:smallest_degree}} \label{sec:appendix_A}

\subsubsection{For $\alpha_1 = 0$}

Since $r_1 \left( \mu_{p_1}, \cdots, \mu_{p_K} \right)$ is a product
of two polynomials as shown in (\ref{eq:function_r1}), the smallest
degree of $r_1 \left( \mu_{p_1}, \cdots, \mu_{p_K} \right)$ is the
sum of the smallest degrees of each polynomial $g \left( \mu_{p_1},
\cdots, \mu_{p_K} \right)$ and $h_1 \left( \mu_{p_1}, \cdots,
\mu_{p_K} \right)$. Let's define $D_{g, smallest}$ as the smallest
degree of the polynomial $g \left( \mu_{p_1}, \cdots, \mu_{p_K}
\right)$. It is easily found that all of the terms in
(\ref{eq:function_g}) have the same degree. Therefore,
\begin{align}
D_{g,smallest} = K(X-Y)+K(K-1) \label{eq:D_g_smallest}
\end{align}
where the degree of $K(X-Y)$ is contributed by the $K$ factors of
the form $\mu_{p_k}^{X-Y}$, and $K(K-1)$ comes from the
$\binom{K}{2}$ factors in the form of $\left( \mu_{p_k} - \mu_{p_j}
\right)^2$.

To calculate the smallest degree of the polynomial $h_1 \left(
\mu_{p_1}, \cdots, \mu_{p_K} \right)$, we need to know the degree of
the polynomial $\frac{\psi\left( \mu_{1}, \cdots, \mu_{Y}
\right)}{g\left( \mu_{p_1}, \cdots, \mu_{p_K} \right)}$. The
polynomial $\psi\left( \mu_{1}, \cdots, \mu_{Y} \right)$ in
(\ref{eq:Polynomial_p}) has $Y$ factors of the form $\mu_i^{X-Y}$
and $\binom{Y}{2}$ factors of the form $\left( \mu_i - \mu_j
\right)^2$. The division by $g\left( \mu_{p_1}, \cdots, \mu_{p_K}
\right)$ makes the common factors eliminated, leaving $(Y-K)$
factors of the form $\mu_i^{X-Y}$ and
\mbox{$( \binom{Y}{2} - \binom{K}{2} )$} factors of the form
$\left( \mu_i - \mu_j \right)^2$.
Hence, the resulting polynomial $\frac{\psi\left( \mu_{1}, \cdots,
\mu_{Y} \right)}{g\left( \mu_{p_1}, \cdots, \mu_{p_K} \right)}$ has
degree
\begin{align}
D_{h_1,org} = (Y-K)(X-Y)+Y(Y-1)-K(K-1) \label{eq:D_h1}
\end{align}
for all of the terms of the polynomial.

Obviously, there exists an integer $\epsilon$ such that $s_{\epsilon}
< p_1 <s_{\epsilon+1}$ since $p_1 > 1$. The integration over $\mu_i$
for $1 \leq i \leq s_{\epsilon}$ in
(\ref{eq:altMarginal_PDF_decomposed_1}) makes these variables vanish
because of the integration to infinity due to $\mu_1$, while that
over the other $\mu_j$ for $s_{\epsilon+1} \leq j \leq s_{(Y-K)}$
converts those variables into the variables $\mu_{p_1}, \cdots,
\mu_{p_K}$. In the meanwhile, all the terms in $\frac{\psi\left(
\mu_{1}, \cdots, \mu_{Y} \right)}{g\left( \mu_{p_1}, \cdots,
\mu_{p_K} \right)}$ have different distributions on the degrees of
the individual variables although they have the same degree as an
entire term. Therefore, the smallest degree of $h_1 \left(
\mu_{p_1}, \cdots, \mu_{p_K} \right)$ is determined by the term
which has the largest degree of those vanishing variables of
$\frac{\psi\left( \mu_{1}, \cdots, \mu_{Y} \right)}{g\left(
\mu_{p_1}, \cdots, \mu_{p_K} \right)}$. It is not necessary to find
all the terms with the largest degree of the vanishing variables.
Instead, we can see that one of those terms, whose degree is
$D_{h_1,org}$, includes the factors
\begin{align}
\prod\limits_{i=1}^{p_1-1} \mu_{i}^{X-Y} \prod\limits_{j>i}^{Y}
\mu_i^2. \label{eq:largest_term}
\end{align}
In this case, the degree corresponding to the vanishing variables in
(\ref{eq:largest_term}) is
\ifCLASSOPTIONonecolumn
\begin{align}
D_{h_1,vanishing} = (p_1-1)(X-Y)+2Y(p_1-1)-p_1(p_1-1)
\label{eq:D_vanished}
\end{align}
\else
\begin{multline}
D_{h_1,vanishing} = (p_1-1)(X-Y) \\
+2Y(p_1-1)-p_1(p_1-1) \label{eq:D_vanished}
\end{multline}
\fi where $(p_1-1)(X-Y)$ is contributed by the $(p_1-1)$ factors of
the form $\mu_i^{X-Y}$, and the rest of the degrees are calculated
from the factors of the form $\mu_i^2$.

Finally, the integration over $\mu_j$ for $s_{\epsilon+1} \leq j
\leq s_{(Y-K)}$ accumulates the degree of the current variables as
well as the previous variables belonging to $\mathbf{s}$. Hence, the
degree of the variables $\mu_k$ for $k \geq p_1$ is kept during the
multiple integration. In addition, during each integration of
$\mu_j$ for $s_{\epsilon+1} \leq j \leq s_{(Y-K)}$, the degree
increases by $1$ due to the fact that $\int_0^{\mu_i} \mu_{i+1}^n
d\mu_{i+1} = \mu_{i}^{n+1}/(n+1)$. Since $s_{\epsilon}$ variables
from the original $(Y-K)$ variables of integration vanished in $h_1
\left( \mu_{p_1}, \cdots, \mu_{p_K} \right)$, the degree to be added
is
\begin{align}
D_{h_1,added} = Y-K-p_1+1 \label{eq:D_added}
\end{align}
where $s_{\epsilon}$ is replaced by $(p_1-1)$. The smallest degree
of \mbox{$r_1 \left( \mu_{p_1}, \cdots, \mu_{p_K} \right)$} can be calculated as
\ifCLASSOPTIONonecolumn
\begin{align}
D_{r_1,smallest} &= D_{g,smallest} + D_{h_1,smallest} 
= D_{g,smallest} + \left(D_{h_1,org} - D_{h_1,vanishing} + D_{h_1,added} \right) \nonumber \\
&=(X-p_1+1)(Y-p_1+1) - K \label{eq:D_smallest}
\end{align}
\else
\begin{align}
D_{r_1,smallest} &= D_{g,smallest} + D_{h_1,smallest} \nonumber \\
&= D_{g,smallest} \nonumber \\
& \quad + \left(D_{h_1,org} - D_{h_1,vanishing} + D_{h_1,added} \right) \nonumber \\
&=(X-p_1+1)(Y-p_1+1) - K \label{eq:D_smallest}
\end{align}
\fi where $D_{h_1,org} - D_{h_1,vanishing}$ stands for the degree
which is kept during the integration over $\mu_j$ for
$s_{\epsilon+1} \leq j \leq s_{(Y-K)}$. If $X$ is equal to $N$, then
$Y$ is $M$, and vice versa, the smallest degree is
$(N-p_1+1)(M-p_1+1) - K$.

\subsubsection{For $\alpha_1 > 0$}

Similarly to the case of $\alpha_1 = 0$, the smallest degree of $r_2
\left( \mu_{p_1}, \cdots, \mu_{p_K} \right)$ can be calculated in
the same manner as in (\ref{eq:D_smallest}). The smallest degrees of
(\ref{eq:D_g_smallest}) and (\ref{eq:D_h1}) apply to this case as
well. However, since $p_1 = 1$ in this case, leading to $p_1 < s_i$
for $i = 1, \cdots, (Y-K)$, no variable vanishes, resulting in
$D_{h_2,vanishing} = 0$. For the same reason, each of the $(Y-K)$
variables of integration adds one degree. Therefore, $D_{h_2,added}
= Y-K$. By writing an equation similar to (\ref{eq:D_smallest}), we
get the smallest degree of the polynomial $r_2\left( \mu_{p_1},
\cdots, \mu_{p_K} \right)$ as
\begin{align}
D_{r_2,smallest} = XY - K. \label{eq:D_r2_smallest}
\end{align}
In general, the smallest degree of the polynomial $r_1\left(
\mu_{p_1}, \cdots, \mu_{p_K} \right)$ or $r_2\left( \mu_{p_1},
\cdots, \mu_{p_K} \right)$ can be expressed as $(N-p_1+1)(M-p_1+1) -
K$ since this holds true even for the case of $\alpha_1 > 0$ where
$p_1 = 1$.

\vspace{-4mm}
\section{Proof of Theorem \ref{theorem:Elemental_Integral}} \label{sec:appendix_B}

The first integral for the variable $\theta_K$ in
(\ref{eq:Exponent_calculation}) can be calculated as
\ifCLASSOPTIONonecolumn
\begin{equation}
\int_0^{\theta_{K-1}} \theta_K^{\beta_K} e^{-\omega \theta_K }
d \theta_K 
= - \sum\limits_{i=1}^{\beta_K + 1} \frac{\beta_K!}{(\beta_K - i +
1)!} \omega^{-i} \theta_{K-1}^{\beta_K + 1 -i} e^{-\omega \theta_{K-1}}
+ \beta_K! \omega^{-(\beta_K + 1)}.
\label{eq:firstintegral}
\end{equation}
\else
\begin{align}
\int_0^{\theta_{K-1}} &\theta_K^{\beta_K} e^{-\omega \theta_K }
d \theta_K \nonumber \\
&= - \sum\limits_{i=1}^{\beta_K + 1} \frac{\beta_K!}{(\beta_K - i +
1)!} \omega^{-i} \theta_{K-1}^{\beta_K + 1 -i} e^{-\omega \theta_{K-1}} \label{eq:firstintegral} \\
&\qquad + \beta_K! \omega^{-(\beta_K + 1)}.\nonumber
\end{align}
\fi
We will ignore all the constants for the simple expression since we
are interested in the exponent of $\omega$. The second integral can
be calculated as
\ifCLASSOPTIONonecolumn
\begin{align}
\int_0^{\theta_{K-2}} \int_0^{\theta_{K-1}}
\theta_{K-1}^{\beta_{K-1}} \theta_K^{\beta_K} e^{-\omega
(\theta_{K-1} + \theta_K) }
& d \theta_K d \theta_{K-1}  = \sum_i \sum_j \omega^{-(i+j)} \theta_{K-2}^{(\beta_K +
\beta_{K-1}+2-i-j)} e^{-2 \omega \theta_{K-2}} \nonumber \\
&\qquad
- \sum_j \omega^{-(\beta_K+1+j)} \theta_{K-2}^{\beta_{K-1}+1-j} e^{-\omega \theta_{K-2}}
+ \omega^{-(\beta_K + \beta_{K-1}+ 2)}. \label{eq:secondintegral}
\end{align}
\else
\begin{align}
\int_0^{\theta_{K-2}} &\int_0^{\theta_{K-1}}
\theta_{K-1}^{\beta_{K-1}} \theta_K^{\beta_K} e^{-\omega
(\theta_{K-1} + \theta_K) }
d \theta_K d \theta_{K-1} \nonumber \\
&= \sum_i \sum_j \omega^{-(i+j)} \theta_{K-2}^{(\beta_K +
\beta_{K-1}+2-i-j)} e^{-2 \omega \theta_{K-2}} \nonumber\\
&\qquad - \sum_j \omega^{-(\beta_K+1+j)} \theta_{K-2}^{\beta_{K-1}+1-j} e^{-\omega \theta_{K-2}} \label{eq:secondintegral} \\
&\qquad + \omega^{-(\beta_K + \beta_{K-1}+ 2)}.\nonumber
\end{align}
\fi
Even though the exact expression for the $k^{th}$ integral can be
obtained by extending the procedure above, it is too complicated. A
simpler method to calculate the exponent of $\omega^{-1}$ after the
final integral is reached by observing the fact that the sum of the
exponent of $\omega^{-1}$ and $\theta_{K-1}$ in
(\ref{eq:firstintegral}) is $(\beta_K + 1)$ for any term. This fact
can also be observed in (\ref{eq:secondintegral}) as $(\beta_K +
\beta_{K-1} + 2)$. By defining $\Gamma_k$ for the $k^{th}$ integral
in the same manner above, we can generalize $\Gamma_k$ as
\begin{align}
\Gamma_k = \sum\limits_{i=1}^k \beta_{K-i+1} + k. \label{eq:Gamma_k}
\end{align}
The final integration is calculated by
\begin{align}
\int_0^{\infty} \theta_1^{\beta_1} \, \xi \left( \omega, \theta_1
\right) e^{-\omega \theta_1} \, d \theta_1
\label{eq:final_integration}
\end{align}
where $\xi(\omega, \theta_1)$ is the $(K-1)^{th}$ integral of the
variables $\omega$ and $\theta_1$ with $\Gamma_{K-1}$. In other
words, (\ref{eq:final_integration}) is the sum of the many terms
which have the form
\begin{align}
\int_0^{\infty} \omega^{-x} \theta_1^{(y + \beta_1)} e^{-z \omega
\theta_1} \, d \theta_1 \label{eq:decomposed_final_integration}
\end{align}
where $x + y = \Gamma_{K-1}$, and $z$ is a constant depending on the
term. We can easily see that (\ref{eq:decomposed_final_integration})
results in
\begin{align}
\delta \omega^{-(x + y + \beta_1 + 1)}
\label{result_final_integration}
\end{align}
where $\delta$ is a constant, and the exponent of $\omega^{-1}$ is
\begin{align}
x + y + \beta_1 + 1 = K + \sum_{i=1}^K \beta_i.
\label{exponent_omega}
\end{align}

\bibliographystyle{mybibstyle}
\bibliography{IEEEabrv,bibliography/journals,bibliography/books,bibliography/conferences,bibliography/thesis}

\end{document}